\documentclass[12pt]{article}
\pdfoutput=1
\usepackage{amssymb}
\usepackage{graphicx}
\usepackage{hyperref}
\usepackage{bm}
\usepackage{color}

\ifx\affiliation\undefined 
\def\affiliation#1{\date{\normalsize #1\\ \today}}
\usepackage[margin=1in]{geometry}
\usepackage[sort&compress,numbers]{natbib}
\else 
\fi

\def\x{{\bm x}}

\def\hp{\hat\phi}
\def\tf{\tilde f}
\def\tF{\tilde F}

\def\sech{\,{\rm sech}}
\title{Electron phase-space hole transverse instability at high magnetic field}
\author{I H Hutchinson}
\affiliation{Plasma Science and Fusion Center, MIT, Cambridge, MA, USA}

\begin{document}

\maketitle

\begin{abstract}
  Analytic treatment is presented of the electrostatic instability of
  an initially planar electron hole in a plasma of effectively
  infinite particle magnetization. It is shown that there is an
  unstable mode consisting of a rigid shift of the hole in the
  trapping direction. Its low frequency is determined by the real part
  of the force balance between the Maxwell stress arising from the
  tranverse wavenumber $k$ and the kinematic jetting from the hole's
  acceleration. The very low growth rate arises from a delicate
  balance in the imaginary part of the force between the
  passing-particle jetting, which is destabilizing, and the resonant
  response of the trapped particles, which is stabilizing. Nearly
  universal scalings of the complex frequency and $k$ with hole depth
  are derived.  Particle in cell simulations show that the
  slow-growing instabilities previously investigated as coupled
  hole-wave phenomena occur at the predicted frequency, but with
  growth rates 2 to 4 times greater than the analytic prediction. This
  higher rate may be caused by a reduced resonant stabilization
  because of numerical phase-space diffusion in the simulations.
\end{abstract}

\section{Introduction}
Long-lived solitary electrostatic structures that are isolated peaks of
positive potential at Debye-length scale, are now routinely observed
as a major high-frequency component of space-plasma
turbulence \citep{Matsumoto1994,Ergun1998,Bale1998,Mangeney1999,Pickett2008,Andersson2009,Wilson2010,Malaspina2013,Malaspina2014,Vasko2015,Mozer2016,Hutchinson2018b,Mozer2018}. They
give rise to what used to be called Broadband Electrostatic Noise that
occurs widely wherever unstable parallel electron distributions are
present; and they are now interpreted as mostly ``electron holes'': a
type of nonlinear Bernstein, Greene, and Kruskal (BGK)
mode \citep{Bernstein1957} in which the potential is self-consistently
maintained by a deficit of electron phase-space density on trapped
orbits \citep{Turikov1984,Schamel1986a,Eliasson2006,Hutchinson2017}. Although
electron holes are routinely produced as the endpoint of
Penrose-unstable one-dimensional Vlasov-Poisson (particle-in-cell or
continuum) kinetic
simulations \citep{Morse1969,Berk1970,Hutchinson2017}, multi-dimensional
simulations observe them to be long-lived only when a magnetic field
in the trapping direction suppresses ``transverse instabilities'' that
tend to break up the electron
holes \citep{Mottez1997,Miyake1998a,Goldman1999,Oppenheim1999,Muschietti2000,Oppenheim2001b,Singh2001,Lu2008}. Even
with a very strong magnetic field, many simulations have observed much
slower growing transverse
instabilities \citep{Goldman1999,Oppenheim1999,Newman2001a,Oppenheim2001b,Umeda2006,Lu2008,Wu2010},
usually associated with coupled long-parallel-wavelength potential
waves, well outside the hole, that are called ``streaks'' or
``whistlers''. These instabilities are important because they may
decide the long-term fate of electron holes in the high-field
regime, causing planar holes to break up into three-dimensional shapes
of limited transverse extent.

This paper presents the first satisfactory theoretical analysis of
transverse electron hole instability in the high-magnetic field
regime. Prior analytical investigations in this regime have
concentrated on coupling to the waves. The pioneering work of Newman
et al. \citep{Newman2001a} correctly identified the importance of kink
oscillation of the hole and calculated its real frequency in a
simplified waterbag model, in agreement with simulation. The external
waves are actually magnetized Langmuir oscillations at the
high-frequency end of the whistler branch of the cold plasma
dispersion relation at substantially oblique propagation:
$\omega\simeq\omega_pk_\perp/k$ ; the other, lower frequency, end of
the branch, at near-parallel propagation, is where the ionospheric
whistler phenomena occur.  However, Newman et al's instability
mechanism was taken to be coupling between the hole and the external
waves. And their analysis inappropriately represented the hole
coupling through a single long-wavelength traveling wave Fourier
mode. That is not what is observed in subsequent simulations, and
cannot rightly represent the localized electron hole's effect on the
wave, which gives a local impulse, a standing wave with a local step
in potential, and is proportional to the hole's acceleration, not just
its displacement. Moreover the wave's effect on the hole is not just
its lowest Fourier mode. Therefore, although their simulations showed
instability, their analysis was based on unjustified coupling
assumptions. The other early published attempt, by Vetoulis and
Oppenheim \citep{Vetoulis2001}, at an analytic understanding of the
high-field instability correctly indentified particle bounce resonance
as an important ingredient, and gave an expression for the resulting
electron distribution function perturbation. However it then took the
perturbing potential to be a single long-wavelength mode, presumably
to represent the wave, and discarded the far more important
perturbation arising from the hole position shift. In other words, the
particle kinetics of the hole was taken to be a small
perturbation to the wave, even in the hole vicinity, rather than the
wave to be a small perturbation to the hole, which is more
appropriate. Berthomier et al \citep{Berthomier2002}, motivated by
measurements of Auroral phenomena, gave a very helpful review of
experiment and theory, and used a different route to solving the
Vlasov-Poisson system but made essentially the same erroneous
assumption that the perturbing potential was purely the wave. The
shortcomings of these competing prior analyses have left the
high-field instability mechanism so far unresolved, even though
simulations continue to observe it. The present paper is aimed at a
rigorous treatment to resolve this uncertainty.

\begin{figure}
  \centering
  \includegraphics[width=.6\hsize]{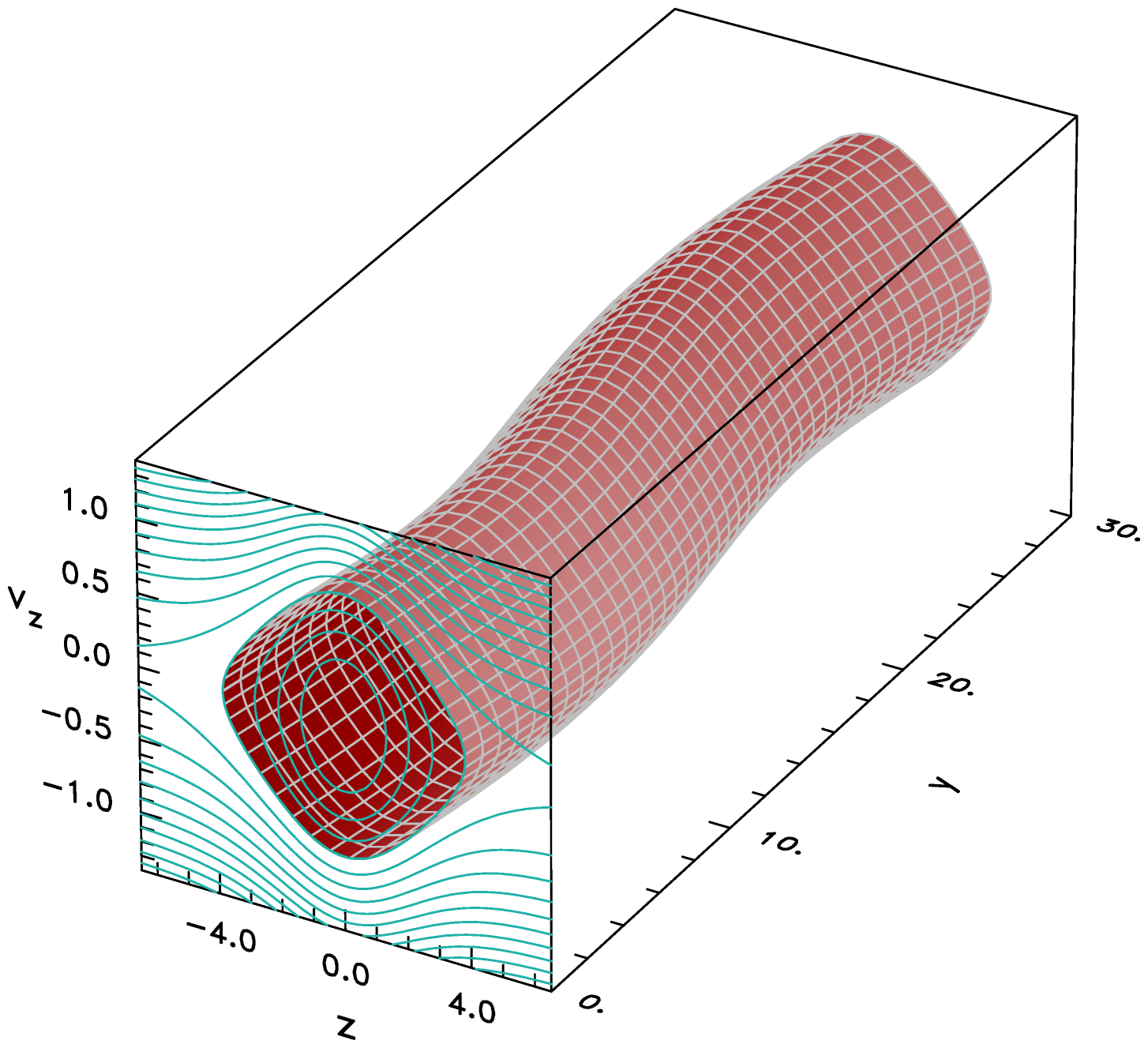}
  \caption{Illustration of the shift kink of an electron hole showing
    phase-space $z,v_z$ contours of constant energy with a specific
    (trapped particle) iso-energy surface rendered as a function of
    transverse position $y$, in three dimensions.}
  \label{fig:isoenergy}
\end{figure}

It analyzes, for an initially planar hole, a perturbation analogous to
a kink of a cylindrical plasma: a uniform shift displacement of the
hole in the direction $z$ of particle trapping with a finite
transverse wave-vector $k_y=k$, as illustrated in Fig.\ \ref{fig:isoenergy}.
This eigenmode structure is adopted as
an ansatz that is well theoretically justified at low frequency; and
has been shown to give accurate values of real and imaginary
frequencies for the transverse instability at low and moderate
magnetic fields. It predicts that the low-field instability is purely
growing \citep{Hutchinson2018,Hutchinson2018a}, that it is stabilized at
a certain B-threshold \citep{Hutchinson2018a}, and is replaced by an
oscillatory instability which then stabilizes above a second
threshold \citep{Hutchinson2019}, all of which are in good quantitative
agreement with PIC simulations. Those simulations, however, like
earlier ones, sometimes observe a residual high-magnetic field oscillatory kink
instability, coupled to external waves, with a much lower growth
rate. It persists to apparently arbitrarily high magnetic field (and
hence cannot be attributable to cyclotron
resonances \citep{Jovanovic2002} since its frequency shows no B-field
dependence), and is the motivating observation behind the present
extension of the analysis.

Suppression of transverse particle motion by the high magnetic field
is a significant simplification, and allows one to derive the
eigenmode equations from elementary one-dimensional understanding of
the Vlasov equation, and to develop purely analytic approximations for
the hole force terms whose balance gives the eigenvalue and hence the
frequency and growth rate. Section 2 explains the derivation and force
balance, and then provides some motivating and explanatory
observations of the important physics, based on numerical orbit
integration, which lays the groundwork for the analytical
solution. Section 3 performs the analysis of the three dominant
imaginary contributions to the particle force arising from a hole
shift perturbation, using well-motivated analytic approximations of
the anharmonic motion of the trapped and passing particles. Together these
determine (in section 4) a universal dispersion relation between the
real and imaginary parts of the frequency $\omega$ and the
corresponding transverse wave-number $k$. Section 5 compares the
results with some particle in cell simulations, and makes the case
that the instability mechanism analysed is probably responsible
for the hole-wave coupled instability observed in simulations, albeit
with some caveats.

The stability analysis takes no account of any induced waves external
to the hole. It shows that there is essentially always a slow-growing
residual transverse instability of a slab hole at high magnetic field,
regardless of the precise field magnitude, and regardless of any
coupling to external waves. This instability might generate
waves \citep{Singh2001a} and the simulations give evidence that it is
influenced and possibly sometimes suppressed or enhanced by hole-wave
coupling, but hole-wave coupling should probably be regarded as a feature, not
an intrinsic causative mechanism, of the instability.

Ions are taken as a uniform immobile background, only electron
dynamics are included, and the external background distribution of
untrapped electrons $f_0$ is taken to be an unshifted Maxwellian. These
simplifications well represent holes that move much faster than the
ion sound speed but slower than the electron thermal speed.
Throughout this paper dimensionless units are used with length
normalized to Debye length $\lambda_D=\sqrt{\epsilon_0T_e/n_0e^2}$,
velocity to electron thermal speeds $v_t=\sqrt{T_e/m_e}$, electric
potential to thermal energy $T_e/e$ and frequency to plasma frequency
$\omega_p=v_t/\lambda_D$ (time normalized to $\omega_p^{-1}$). In
these normalized units the electron mass ($m_e$) is unity and is
omitted from the equations but normalized electron charge is $q_e=-1$,
and is retained. The parallel energy $W$ is written
${1\over 2}v_z^2+q_e\phi$.

\section{High-B Instability}

\subsection{The Vlasov Poisson system}

A rigorous analysis of the problem of transverse instability at
arbitrary magnetic field strength has been presented in
 \citep{Hutchinson2018a}, which should be consulted for general
mathematical detail. We proceed more simply here by making the
early approximation of high magnetic field. A linearized analytic
treatment of electrostatic instability of a magnetized electron hole
depends on the first order perturbation to the distribution function
$f_1$ caused by a potential perturbation $\phi_1$. It is found by
integrating Vlasov's equation along the equilibrium (zeroth order)
helical orbits, which are the equation's ``characteristics''. For a
collisionless situation $f$ is constant along orbits.

The integration can be expressed as an expansion in harmonics of the
cyclotron frequency ($\Omega=eB/m_e$). However, if the magnetic field
is strong enough, only the $m=0$ harmonic is important.  Physically,
this high-field approximation amounts to accounting only for
particles' motion along the magnetic field, and ignoring cross-field
motion, taking the Larmor radius (and cross-field drifts) to be
negligibly small. Vlasov's equation is then essentially
one-dimensional so we need only consider the parallel velocity
distribution, denoted $f(v)$.

Because the equilibrium is non-uniform in the (trapping)
$z$-direction, uncoupled Fourier representation of the potential
variation is possible only for the transverse direction (taken as $y$
without loss of generality) orthogonal to $z$.  The $z$-dependence of the
linearized eigenmode must be expressed in a full-wave manner by writing
\begin{equation}
  \label{eq:eigenmode}
  \phi_1(\x,t)=\hp(z)\exp i(ky-\omega t).
\end{equation}

One way to derive the solution of Vlasov's equation intuitively is to
recognize that if a small \emph{time-independent} potential
perturbation is applied, then particles' (parallel) energy is still
conserved as they approach from far past time (and
distance). Consequently the perturbation to the distribution function
($f_1$) at fixed velocity arises purely as a result of the
perturbation to the potential in the form
$f_{ 1}(z) = q_e\phi_1(z){\partial f_{0}\over\partial W}$. This
equation expresses the conservation of distribution function along
constant energy orbits and the fact that the potential perturbation
causes an orbit at fixed velocity to correspond to an energy different
(in the distant past, where the distribution is $f_0$) by
$q_e\phi_1$. This component is commonly referred to as the
``adiabatic'' perturbation.

However when the perturbation is \emph{time-dependent} an additional
effect of particle energization occurs. Particles no longer move with
constant energy. Instead their energy has an instantaneous rate of
increase equal to $q_e{\partial \phi_1\over \partial t}$, at every
position in the past orbit. The energy change from the distant past
can be written
${\cal E=}\int_{-\infty}^t q_e{\partial\over \partial
  t}[\phi_1(z(\tau),\tau)] d\tau$.
The starting $f$ value (still conserved along the perturbed orbit)
thus corresponds to energy smaller by ${\cal E}$.  Therefore the
distribution perturbation acquires a second ``non-adiabatic''
component $-{\cal E}{\partial f_0\over\partial W}$ that, for harmonic
time dependence $\propto {\rm e}^{-i\omega t}$, gives a total
\begin{eqnarray}\label{eq:f1magnetic}
  f_{ 1}=  
  q_e\phi_1(t){\partial f_{0}\over\partial W}
  +
  q_e i\omega\Phi {\rm e}^{i(ky-\omega t)}
  {\partial f_{0}\over \partial W},
\end{eqnarray}
where
\begin{equation}
  \label{eq:phim}
  \Phi (z,t)\equiv 
  \int_{-\infty}^t \hp(z(\tau)){\rm e}^{-i\omega(\tau-t)}d\tau,
\end{equation}
and $z(\tau)=z(t)+\int_t^\tau v_z(t')dt'$ is the position at earlier
time $\tau$ (see  \citep{Hutchinson2018a} eq:5.6).  For positive
imaginary part of $\omega$ ($\omega_i>0$) the integral converges. We
denote the second term of eq.\ (\ref{eq:f1magnetic}) omitting the
dependence ${\rm e}^{i(ky-\omega t)}$, as
$\tf\equiv q_ei\omega\Phi \partial f_0/\partial W$, which is the
``non-adiabatic'' distribution perturbation.

The main formal difficulty is to find the shape of the eigenfunction $\hp(z)$
which self-consistently satisfies the perturbed Poisson equation: an
integro-differential eigenproblem. For slow time dependence relative
to particle transit time, it can be argued on general grounds that the
eigenmode consists of a spatial shift (by small distance $\xi$
independent of position) of the equilibrium potential profile
($\phi_0 (z)$) (see  \citep{Hutchinson2018a} section 3.1) giving:
\begin{equation}
  \label{eq:shiftmode}
  \hp = - \xi {\partial \phi_0 \over \partial z}.
\end{equation}
The frequencies we care about have periods not much longer than the
particle transit time, at least for particles with total energy near
zero; so this shift form cannot be expected to hold exactly. However,
we can obtain a good approximation to the corresponding eigenvalue of
our system by expressing it in terms of a ``Rayleigh Quotient''.  This
mathematical procedure is equivalent to requiring the conservation of
total $z$-momentum under the influence of the assumed shift eigenmode.
(See \citep{Hutchinson2018a} section 3, and
\citep{Hutchinson2016,Zhou2017}).  This amounts to a
``kinematic'' treatment of the hole as a composite object.  

The $z$-momentum balance can be derived in an elementary way by
applying the zeroth and first order Poisson equations
${d^2\phi_0/dz^2}=-{\rho_0/\epsilon_0}$, and
${d^2\phi_1/ dz^2}-k^2\phi_1=-{\rho_1/\epsilon_0}$ (where
$\rho$ is the charge density) to the integral expression for the first
order hole force $\int\rho_0(-d\phi_1/dz) dz$, using judicious integrations by
parts as follows:
\begin{eqnarray}
  \label{eq:Forcederiv}
  -\int\rho_0{d\phi_1\over dz} dz 
  &=& \epsilon_0\int{d^2\phi_0\over dz^2}{d\phi_1\over dz}dz
=-\epsilon_0\int{d\phi_0\over dz}{d^2\phi_1\over dz^2}dz\nonumber\\
&=&\int{d\phi_0\over dz}(\rho_1-\epsilon_0 k^2\phi_1)dz.
\end{eqnarray}
Now, since $\rho_0$ is a function of $\phi_0$, we can also integrate by
parts the other way to find
\begin{equation}
  \label{eq:ForceD2}
   -\int\rho_0{d\phi_1\over dz} dz 
   =\int{d\rho_0\over d\phi_0}{d\phi_0\over dz}\phi_1dz.
\end{equation}
Combining and rearranging these expressions we get
\begin{eqnarray}
  \label{eq:forcebalance}
  F_E \equiv -\epsilon_0 k^2 \int {d\phi_0 \over dz}\phi_1 dz 
&=&-\int{d\phi_0 \over dz}\left(\rho_1-{d\rho_0\over d\phi_0}\phi_1\right)dz
\nonumber\\
&=& -\int{d\phi_0 \over dz}\left( \int q_e\tf dv_z\right)   dz
 \equiv   \tF.
\end{eqnarray}
The force $F_E$ consists of transfer by Maxwell stress in the
$y$-direction of $z$-momentum ($d/dy(E_yE_x$)). It acts in a direction
so as to increase the kink amplitude, in a manner analogous to a
compressed spring.  The force $\tF$ is exerted by the equilibrium
potential on the non-adiabatic part of the charge density perturbation
which is the jetting. Its real part is proportional to kink
acceleration, and acts like a negative inertia. The eigenvalue
equation (\ref{eq:forcebalance}) is that they must balance; we
substitute into it the shift-mode form for $\phi_1$, eqs.\
(\ref{eq:shiftmode} and \ref{eq:eigenmode}). To lowest order,
satisfying the real part of the force equation, the
result is an oscillation at a real frequency whose square is
proportional to the ratio of the negative tension effect and the
negative inertia.

Both the real and imaginary parts of the complex momentum balance
equation $F_E=\tF$ must be zero. But $F_E$ is real and positive for
real $k$, and in this high-B approximation $k$ appears only in $F_E$
and not in $\tF$. If we regard $k$ as a free choice, then provided the
sign of $\Re(\tF)$ is positive, we can always satisfy $\Re(\tF-F_E)=0$
by simply choosing the appropriate value for $k$.  Consequently it is
only the imaginary part $\Im(\tF)$ (which is independent of $k$) that
determines whether there exists a solution of the dispersion relation
with a frequency $\omega=\omega_r+i\omega_i$ in the upper half plane
($\omega_i>0$), implying instability.  We denote
contributions from trapped (negative energy, $W<0$) and passing
($W>0$) particles with subscripts `t' and `p' respectively, and write
$\tF=\tF_t+\tF_p$. At frequencies low compared with the transit time
of thermal electrons across the hole, $\Re(\tF_t)>0$, $\Re(\tF_p)<0$,
and $|\Re(\tF_t)|>|\Re(\tF_p)|$ so $\Re(\tF)$ is indeed
positive.

\subsection{Numerical Evaluation Observations}

\begin{figure}
  \centering
  \includegraphics[width=0.48\hsize]{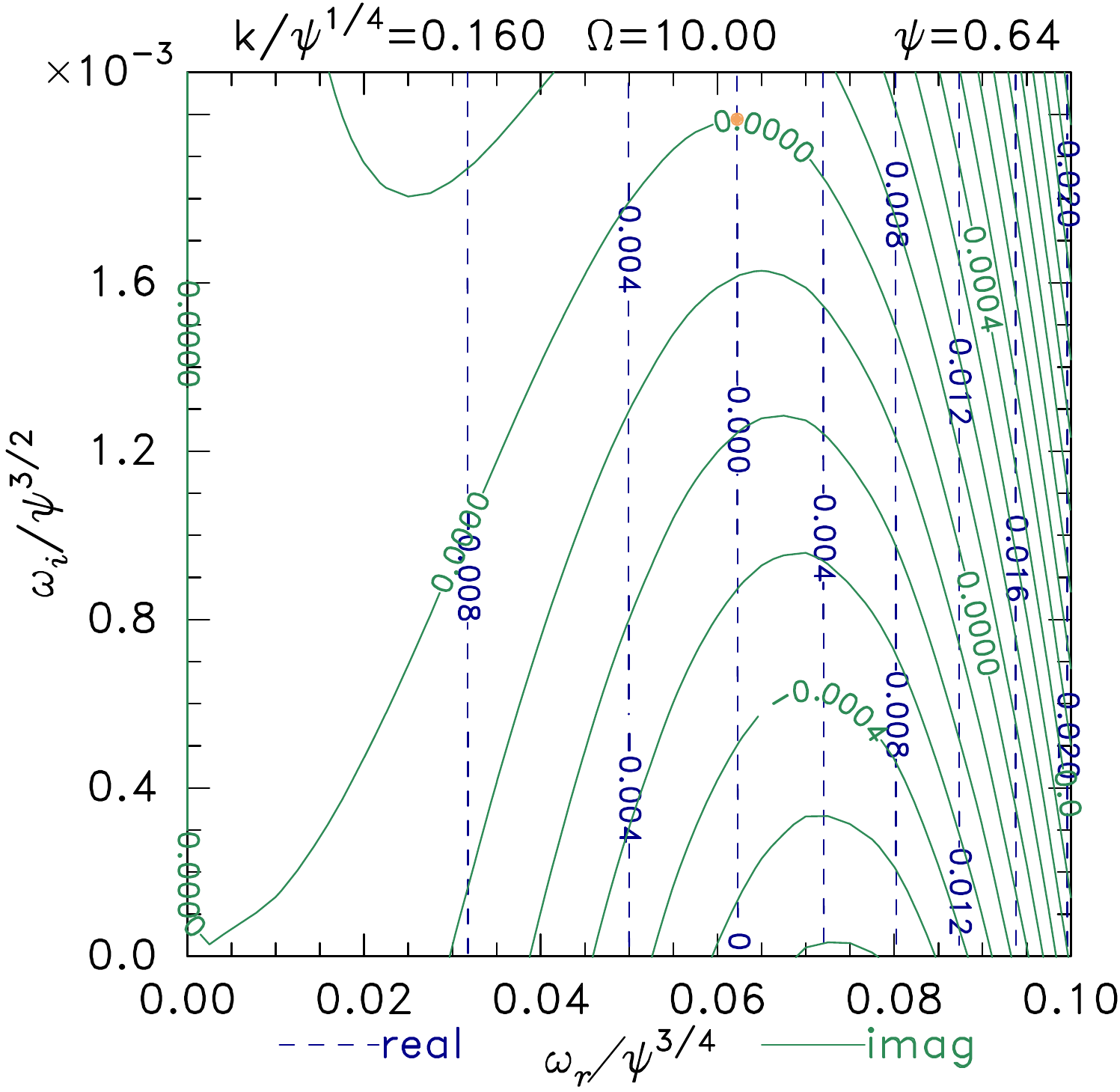}
  \includegraphics[width=0.48\hsize]{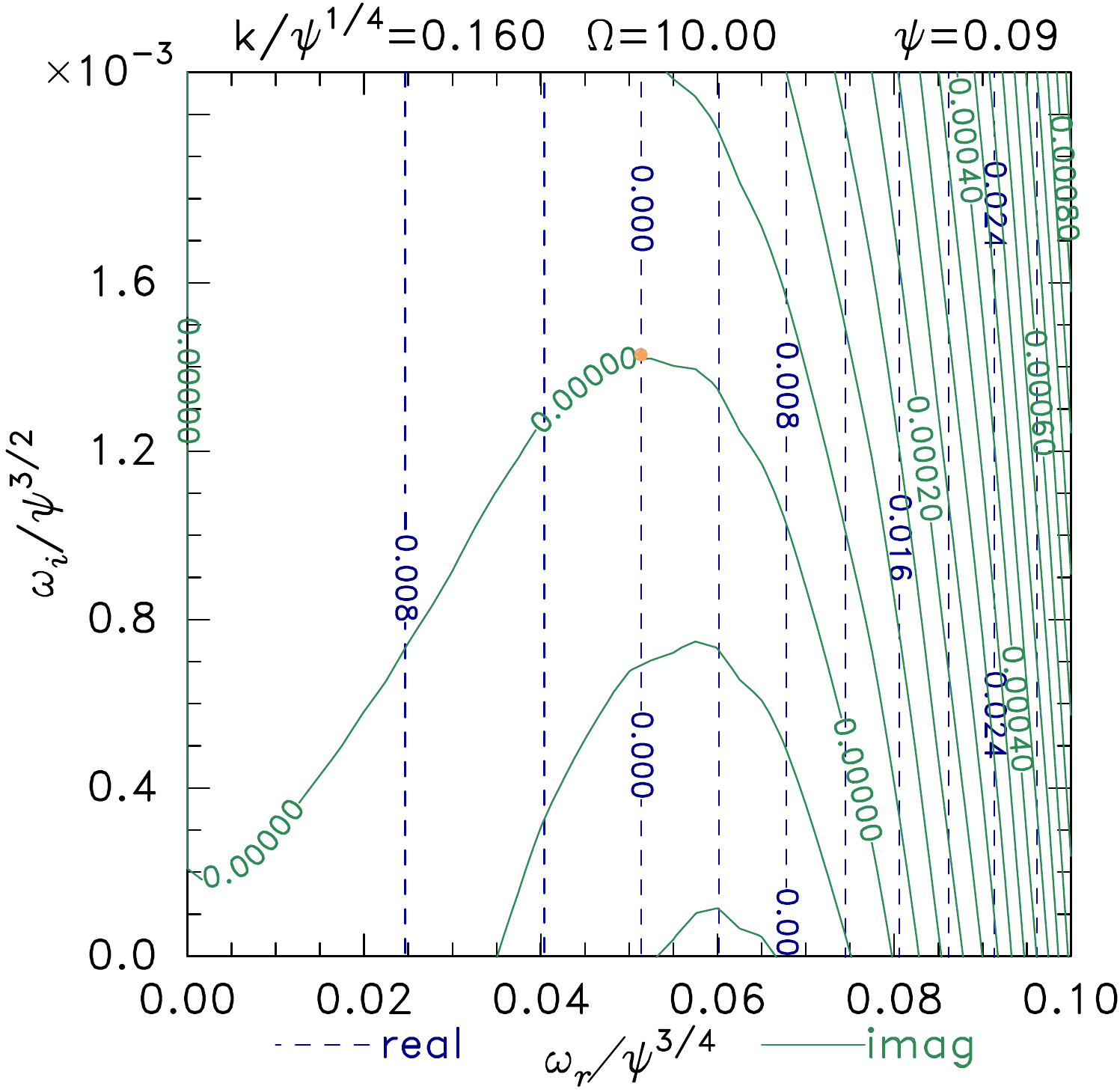}
  \caption{Contours of real and imaginary parts of
    $\tF-F_E$. Instability occurs for different $k$ at different
    places along the zero contour $\Im(\tF-F_E)=0$. For the particular
  $k$ chosen, the eigenfrequency is show as a point.}
  \label{fig:fcontours}
\end{figure}

A numerical implementation of the required integrations to find $\tF$
has previously been developed \citep{Hutchinson2018a} for the specific hole equilibrium
\begin{equation}
\label{holeequil}
  \phi_0(z) = \psi \sech^4(z/4)
\end{equation}
where the constant $\psi$ is the maximum hole potential: the ``depth''
of the hole. This code actually performs the integrations for
arbitrary magnetic field strength but works well for the present
high-B case, with some modifications, described later, newly
implemented to allow accurate evaluations at $\omega_i\to 0$.  It
shows there to be unstable solutions at low frequency as illustrated
by the contours of $\tF$ plotted in Fig \ref{fig:fcontours}. An
unstable mode occurs at the intersection of the zero-contours of the
real and imaginary parts of $\tF$.
Notice that the real part of the frequency is small even for the deep
hole $\psi=0.64$, but the imaginary part is far smaller
$\omega_i\lesssim 0.02\psi^{0.75}\omega_r$. The shape of the contours
is approximately similar for different hole depths, when the
frequencies are scaled to $\omega_r/\psi^{0.75}$,
$\omega_i/\psi^{1.5}$, and the wavenumber to $k/\psi^{0.25}$. The
vertical contours of real part show that there is negligible influence
of $\omega_i$ on $\Re(\tF)$ in this low frequency region. The plots
are for a specific finite field $\Omega=10$, but are essentially
unchanged for any $\Omega\gtrsim 5$, indicating that we are well into
the high-B, one-dimensional motion, regime.  We wish to derive
analytically the shape of these contours in order to identify the
controlling physics of this instability.

We concentrate on the decisive imaginary part of
\begin{eqnarray}
  \label{eq:F}
  \tF&=&-(i\omega)\xi \int q_e{d\phi_0\over dz}\int {\partial
    f_0\over \partial W} q_e\int_{-\infty}^t\hat\phi(z(\tau))
  {\rm e}^{-i\omega(\tau-t)}d\tau dv dz. 
\end{eqnarray}
The resulting lowest order in $\omega$ contribution to $\tF$ is
proportional to $\omega^2$ times a real quantity \citep{Hutchinson2018a}. It therefore
gives rise to an imaginary component
$\Im(\tF)= 2{\omega_i\over\omega_r}\Re(\tF)$. The total contribution
of this type includes both trapped and passing terms but the trapped
real force is typically about five times larger than the passing real
force, and will be our focus in the analytic approximation. However,
Fig.\ \ref{fig:fcontours} shows there are non-zero imaginary
components at $\omega_i\to 0$.  We therefore seek, in addition, the
imaginary component of the trapped particle force $\tF_t$ of lowest
order in $\omega$ that does \emph{not} depend on $\omega_i$. It comes
from accounting to higher order for the variation of the
${\rm e}^{-i\omega (\tau-t)}$ factor, and is contributed by the
resonance between the eigenfrequency $\omega$ and the bounce frequency
$\omega_b$ of some trapped particles. There is also an important imaginary
component of $\tF_p$ that does not depend on $\omega_i$.

\begin{figure}
  \centering
  \includegraphics[width=0.5\hsize]{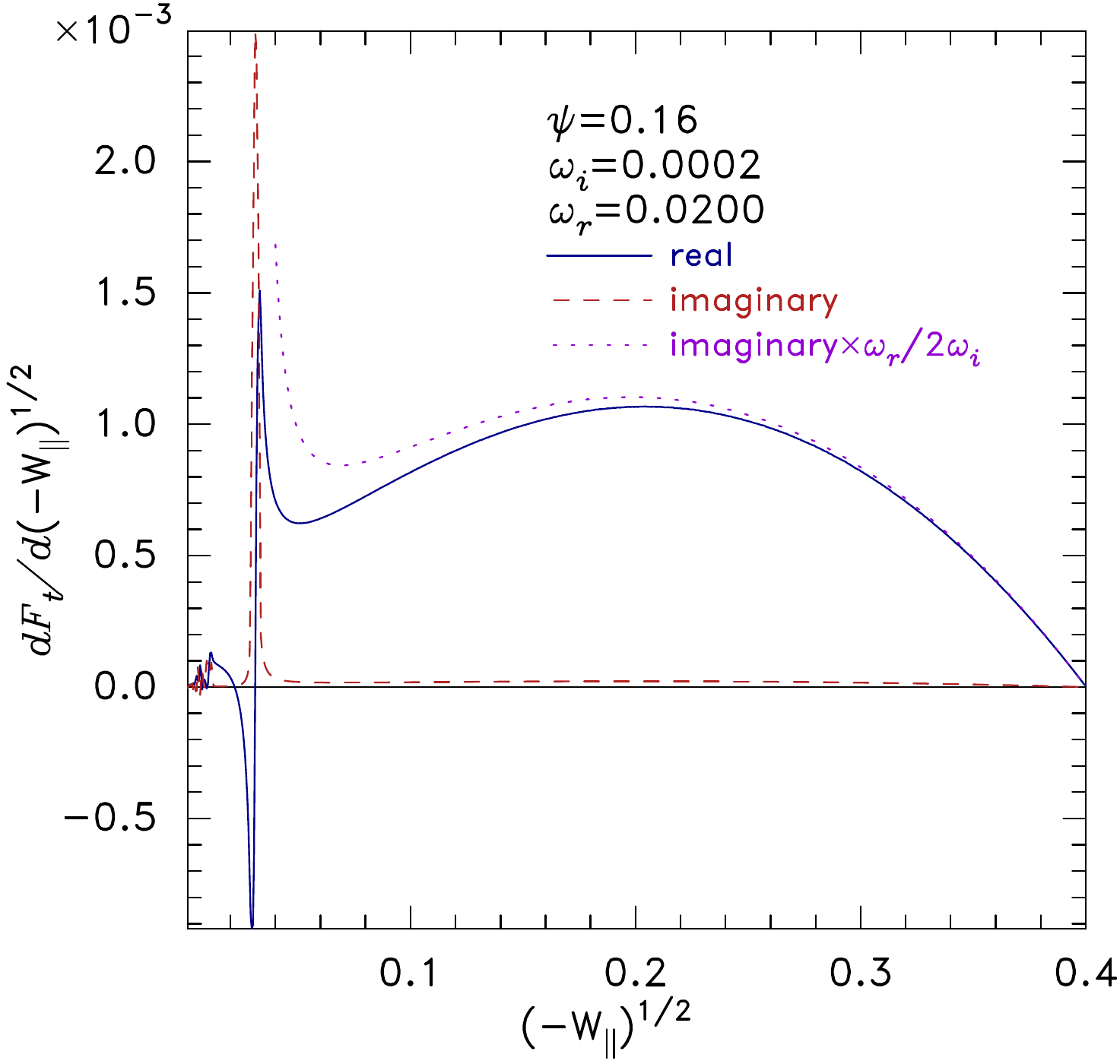}
  \caption{Example of contributions to the trapped particle force from
    different particle energies, showing the resonance between bounce
    frequency and eigenfrequency. Components: Solid line, real; dashed line,
    imaginary; dotted line, $\omega_r/2\omega_i\times$imaginary.}
  \label{fig:resforce}
\end{figure}

One can exchange the order of integration in eq.\ (\ref{eq:F}) so as
to perform the velocity integration last. The notation
${d\tF_t\over du}$ is used to denote the quantity that when integrated
$du$ over any velocity-dependent variable $u$ gives $\tF_t$. Fig.\
\ref{fig:resforce} shows an example of the real and imaginary parts of
${d\tF_t\over d(-W)^{1/2}}$. The area under the curves gives the total
force.
The parameter $(-W)^{1/2}$ runs from zero for the highest energy
(marginally) trapped particles to $\sqrt{\psi}$ for the deepest
trapped particles. As will be shown shortly, it is approximately
proportional to the bounce frequency, because of the anharmonic shape
of the electron hole potential $\phi_0$, which falls exponentially to
zero in the wings.  In Fig. \ref{fig:resforce} we observe for this low
frequency ($\omega_r=0.02$, $\omega_i=2\times10^{-4}$) case a resonant
response at a low value of $(-W)^{1/2}$ (weakly trapped particles)
corresponding to $\omega_r\simeq \omega_b$. It is obvious that in
addition to the substantial real force, arising from the area under
the solid curve, there is a smaller imaginary force that is dominated
by the contribution from the resonance. Resonances also occur at odd
harmonics of the fundamental bounce frequency of particles having
lower $\omega_b$ that are closer to zero energy (the trapped/passing
boundary). However their contribution is much smaller and can be
ignored.

The low-level non-resonant imaginary contribution that extends across
the entire $(-W)^{1/2}$ range in Fig. \ref{fig:resforce} arises from
the component $\Im(\tF)= 2{\omega_i\over\omega_r} \Re(\tF)$ (as is
verified by the dotted line, which equals $\Re(\tF)$ far from resonance).
It tends to zero as $\omega_i\to 0$, as expected from analysis; but
the narrow resonant contribution does not. Instead it becomes narrower
and higher with a convergent total area. This fact poses a numerical
challenge at small $\omega_i$ for the code that calculates $\tF$ by
discrete integration. The challenge is overcome by adopting the approach made
familiar in plasma physics in the context of Landau damping: namely
displacing the contour of integration in the
$\omega_b\leftrightarrow v$ plane so that it remains below the pole at
$\omega_b=\omega$ as $\omega_i$ decreases toward (or even beyond)
zero.  Keeping the integration contour sufficiently below the pole
limits the narrowness and height of the resonance, allowing it to be
numerically resolved. We now obtain approximate analytic expressions
for the important contributions to the imaginary force at low
frequency.

\section{Analytic Treatment of the Hole Momentum}

\subsection{Bounce Orbit Integration}

First we obtain the relationship between the bounce frequency of
trapped particles, $\omega_b$, and (unperturbed) orbit energy $W$. To
approximate analytically for low frequency we observe that orbits near
the separatrix ($W\to 0$) have low $\omega_b$ because they dwell most
of their time near the turning points. The potential and its gradient
are small there, and for the $\sech^4(z/4)$ hole they can be
approximated as
$-{d\phi_0\over d|z|} \simeq \phi_0\simeq \psi 16 {\rm e}^{-|z|}$.
Actually, the mechanism of Debye shielding requires, in the far wing,
that $\phi_0 \propto {\rm e}^{-|z|}$ for any hole shape that does not
have infinite gradients of $f$ at the separatrix \citep{Hutchinson2017}. Therefore the
treatment has much wider application than the specific $\sech^4$
hole-shape. In the wing (where $\phi_0\sim-W$) we may thus take
$d\phi_0/\phi_0\simeq -d|z|$, and can write the relationship
between time $t$ to pass from the turning point to a smaller $|z|$,
corresponding to a larger $\phi_0$, as
\begin{equation}
  \label{eq:timeint}
  t(z) = \int {-d|z|\over |v|}\simeq
  {1\over \sqrt{2}}
  \int {d\phi_0\over \phi_0\sqrt{\phi_0+W}}
  = {1\over\sqrt{2}} {2\over\sqrt{-W}}\tan^{-1}\sqrt{{\phi_0\over(-W)}-1},
\end{equation}
whose inverse is 
\begin{equation}
  \label{eq:timeinv}
  {\phi_0\over -W} = 1 + \tan\left(t \sqrt{-W/ 2}\right).
\end{equation}
A quarter period of the orbit has been reached when $-\phi_0/W$ is
large, which occurs when the value of $t\sqrt{-W/2}$ is
$\simeq\pi/2$. Consequently the orbit period is approximately
$t_b=2\pi /({\sqrt{-W/2}})$, and $\omega_b=\sqrt{-W/2}$.

We shall shortly also need the following average over the orbit, which
can be evaluated for the first quarter period using the relation
(\ref{eq:timeinv}), initially neglecting the distinction between
$\sqrt{-W/2}$ and $\omega_b$:
\begin{eqnarray}
  \label{eq:phiave2}
  \left\langle{\phi_0\over -W}\cos\omega_bt\right\rangle
&\simeq& {1\over t}\int_0^t  \cos\omega_bt' + \sin\omega_bt'\; dt'
=[\sin\omega_bt-\cos\omega_bt+1]/\omega_bt.
\end{eqnarray}
Taking $\omega_bt=\pi/2$, we get an average
equal, by symmetry, to that for a full orbit:
\begin{equation}
  \label{eq:phiaveapx2}
  \left\langle{\phi_0}\sin\omega_b\tau'\right\rangle
  =\left\langle{\phi_0}\cos\omega_bt\right\rangle
  \simeq {4\over \pi}(-W)\simeq {8\over \pi}\omega_b^2;
\end{equation}
($\tau'$ is a shifted time, measured from the center of the orbit:
$\omega_b\tau'=\pi/2+\omega_b t$.)

Although these results are exact as $W\to0$, and over-estimate
$\omega_b$ by only a factor $\sqrt{2}$ at the other extreme
$-W\to \psi$, their inaccuracy will turn out to be numerically significant. It
arises because the estimate of $t_b$ has neglected the extra time it
takes the orbit to pass from the place where the relationship
$d\phi_0/\phi_0\simeq -d|z|$ is broken by the rounded potential peak,
to the exact center of the hole $z=0$.  We track the correction by
writing $\omega_b=A\sqrt{-W/2}$ where $A$ is a correcting factor close
to but slightly below unity. Approximately the same factor applies to
the mapping of $W$ to $\omega_b^2$ in eq.\ (\ref{eq:phiaveapx2}) which
will be written
$\left\langle{\phi_0}\sin\omega_b\tau'\right\rangle\simeq {8\over
  \pi}\omega_b^2/A^2$.

\subsection{Resonant Force}

To evaluate the resonant contribution of trapped particles to the
force, it is convenient to integrate eq.\ (\ref{eq:phim}) twice by
parts using the fact that $\hp=-\xi{d\phi_0\over dz} =\xi{dv\over d\tau}/q_e$:
\begin{equation}
  \label{eq:phimint}
  {q_e\Phi(t)\over\xi}= \int_{-\infty}^t{dv\over d\tau} {\rm e}^{-i\omega(\tau-t)}d\tau
  =v(t)+i\omega z(t)+(i\omega)^2\int_{-\infty}^tz(\tau) {\rm
    e}^{-i\omega(\tau-t)}d\tau. 
\end{equation}
This is exact for an exact shift mode. If the eigenmode deviates from
a shift mode (for example by having a shift $\xi$ that is a
non-uniform function of $|z|$) but retains the properties of the
shiftmode in that it is antisymmetric and localized to the hole, then
the treatment remains valid provided we reinterpret in $\Phi$ the
meaning of $v$ to be $\int q_e\hp d\tau/\xi$, and $z=\int v d\tau$.

The benefit of this reexpression is that, measuring $\tau$ from where
$z=0$, $z(\tau)$ is an approximate square-wave in time on low bounce
frequency orbits. Also, when we integrate the resulting
$\tf {d\phi_0\over d z}$ over the hole extent, to obtain $\tF_t$, the
antisymmetry of ${d\phi_0\over d z}$ annihilates any symmetric part of
$\tf$ and hence of $\Phi$. Therefore the term $v(t)$ can be dropped.
Choose the zero of $\tau$ and $t$ to be where $v$ is positive, during
the bounce period chosen to minimize $|t|$. Then the integral in eq.\
(\ref{eq:phimint}) can be separated into two parts: $\int_0^t$ with
$|t|\le t_b/2$ plus $\int_{-\infty}^0$. Only $\int_{-\infty}^0$
contributes to the resonant term. The other (non-resonant) terms will
be dealt with later.

Represent the square-wave during the final orbit as
$z(t)=sign(t)z_A$. The resonant part of the integral may then be evaluated
as
\begin{eqnarray}
  \label{eq:zreson}
  {q_e\Phi_R\over(i\omega)^2\xi}
 &=&\int_{-\infty}^0z(\tau) {\rm e}^{-i\omega(\tau-t)}d\tau
                     ={1\over 1-{\rm e}^{i\omega t_b}}
  \int_{-t_b}^0 z(\tau)  {\rm e}^{-i\omega(\tau-t)}d\tau\nonumber\\
&=& {z_A{\rm e}^{i\omega t}\over 1-{\rm e}^{i\omega t_b}}{1\over i\omega}\left(
    \left[{\rm e}^{-i\omega\tau}\right]_{-t_b/2}^0
    -\left[{\rm e}^{-i\omega\tau}\right]_{-t_b}^{-t_b/2}
\right)\\
&=&  {z_A{\rm e}^{i\omega t}\over i\omega}
    {\left(1-{\rm e}^{i\omega t_b/2}\right)^2\over 1-{\rm e}^{i\omega t_b}} 
    ={z_A{\rm e}^{i\omega t}\over i\omega}
    {1- {\rm e}^{i\omega t_b/2} \over 1+ {\rm e}^{i\omega t_b/2}}
    \left[={z_A{\rm e}^{i\omega t}\over \omega}\tan(\omega t_b/4)\right].\nonumber
\end{eqnarray}
The denominator $1+ {\rm e}^{i\omega t_b/2}$ becomes zero, giving
resonances, when $\omega t_b=2\pi \ell$ with $\ell$ an odd integer.
In the vicinity of the $\ell$th (odd) resonance
\begin{equation}
  \label{eq:oddresonance}
  {1- {\rm e}^{i\omega t_b/2} \over 1+ {\rm e}^{i\omega t_b/2}}\simeq 
  {2\over 1-{\rm e}^{i\pi(\omega-\ell\omega_b)/\omega_b}}\simeq
  {-2\omega_b\over i\pi(\omega-\ell\omega_b) }.
\end{equation}
Thus at resonance 
\begin{equation}
  \label{eq:tfres}
  \tf_R =q_e i\omega\Phi_R {\partial f_{0}\over \partial W}
  \simeq -\xi {\partial f_0\over \partial W}
  {z_A{\rm e}^{i\omega t}} {(i\omega)^22\omega_b\over i\pi(\omega-\ell\omega_b) },
\end{equation}
in which $t$ represents the end point $z(t)$ (zero when $t=0$) of the
orbit: the position where $\tf$ is being evaluated.  Also, in general,
for small positive imaginary part $\omega_i$, the integral through
(strictly close under) a resonance is
\begin{equation}
  \label{eq:gresint}
  \int_{-}^+ {g(\omega_b) \over (\ell \omega_b-\omega_r-i\omega_i)}
d\omega_b\simeq {i\pi\over \ell} g(\omega_r/\ell)
\end{equation}
for any slowly varying function $g$.

We now need to get the total non-adiabatic force by integrating
$-q_e {d\phi_0\over dz}\tf_R$ over the relevant phase space $dz\, dv$.
Our current interest is the trapped particles, which occupy a finite
area of phase space bounded by the separatrix. The best way to carry
out this area integral is not along fixed values of $v$ or $z$, but
along fixed values of energy, following the orbits along which
particles move as a function of time (denoted $\tau'$). It is
convenient to represent the orbit energy by the value of the velocity
at the hole center $z=0$ where the potential has its single maximum,
$\psi$, because every orbit does in fact pass through this position. We
write this velocity $v_\psi$ (taken positive) so that (the negative
quantity) $W=q_e\psi+{1\over 2}v_\psi^2=q_e\phi(z)+{1\over
  2}v(z)^2$.
Then, since at constant energy $vdv=v_\psi dv_\psi$, we have
$dz\,dv \to vd\tau'\, dv =d\tau' v_\psi dv_\psi= d\tau'dW \simeq
-d\tau'4\omega_bd\omega_b/A^2$
(using $\omega_b^2=-A^2W/2$).  Incidentally, the numerical integration code is
implemented in the same way, but it uses none of the present
approximations.  

The resonant force as $\omega_i\to 0$ can then be written, using first
the parity relations, and then eqs.\ (\ref{eq:tfres}),
(\ref{eq:gresint}), and (\ref{eq:phiaveapx2}), as
\begin{eqnarray}
  \label{eq:taupint}
  \tF_R &=& -q_e\int_0^{\sqrt{-2q_e\psi}} \int_{-t_b/2}^{t_b/2} 
                {d\phi_0\over dz}\tf_R(\tau')
                d\tau'v_\psi dv_\psi\nonumber\\
        &\simeq& -q_e\int_0^{\sqrt{-2q_e\psi}}\int_0^{t_b/2}
             {d\phi_0\over d|z|}[\tf_R(\tau')-\tf_R(-\tau')] d\tau'v_\psi dv_\psi
                     \nonumber\\ 
        &\simeq& -q_e\xi\int^0_{\sqrt{\psi}/2}
                 {\partial f_0\over \partial W}
                {(i\omega)^2z_A 2\omega_b\over i\pi(\omega-\ell\omega_b) }
                \int_0^{t_b/2}\phi_02i\sin(\omega \tau')
                d\tau'4\omega_bd\omega_b/A^2
                \nonumber\\
        &\simeq& q_e\xi\int^{\sqrt{\psi}/2}_0
                 {\partial f_0\over \partial W}
                {(i\omega)^2z_A 2\omega_b\over i\pi(\omega-\ell\omega_b) }
                i\left\langle\phi_0\sin(\omega \tau')\right\rangle
                t_b4\omega_bd\omega_b/A^2
                \nonumber\\
        &\simeq& q_e\xi
                     \left.{\partial f_0\over \partial W}\right|_R
                (i\omega_r)^2z_A{2\omega_R\over \ell}
                {i}\left\langle\phi_0\cos(\omega_r t)\right\rangle
                     8\pi/A^2\nonumber\\
        &\simeq& -iq_e\xi z_A 
                     \left.{\partial f_0\over \partial W}\right|_R
                     128\omega_{R}^5/A^4,       \qquad {\rm for}\ \ell=1.
\end{eqnarray}
where $\omega_{R}=\omega_r/\ell$ is the resonant bounce frequency.
This surprisingly simple expression has not to my knowledge been
previously discovered. 

It should be remarked that eq.\ (\ref{eq:taupint}) has no explicit
dependence on the hole depth $\psi$. However, the (positive) value of
${\partial f_0\over \partial W}$ varies generally as
$\sim\psi^{-1/2}$ \citep{Hutchinson2015,Hutchinson2017}; more specifically, for a
$\sech^4(z/4)$ hole it can be shown analytically that
\begin{eqnarray}
  \label{eq:f0sech4}
  f_0&=&
         f_{0s}\left[{2\over\sqrt{\pi}}\sqrt{-W}
         +{15\over16}\sqrt{\pi\over\psi}W
         +\exp(-W){\rm erfc}(\sqrt{-W})\right],\\
{\partial f_0\over \partial W} &=&
f_{0s}\left[{15\over16}\sqrt{\pi\over\psi}-\exp(-W){\rm erfc}(\sqrt{-W})\right],  
\end{eqnarray}
where $f_{0s}$ is the distribution function at the separatrix ($W=0$,
$\exp(0){\rm erfc}(0)=1$), and for unit density
$f_{0s}=1/\sqrt{2\pi}$.  We will write this
${\partial f_0\over \partial W} \simeq D(15/16)/\sqrt{2\psi}$, where
$D$ is a correction factor approximately equal to $1-(16/15)\sqrt{\psi/\pi}$,
somewhat smaller than unity.  For a general (non-shift) antisymmetric
potential perturbation $\hat\phi$ one should take
$\xi z_A=-\int\int \hat\phi(\tau') d\tau' d\tau$, in accordance with
the remarks following eq.\ (\ref{eq:phimint}). The fifth power
dependence upon $\omega_R$ is one guarantee that $\ell=3$ and higher
harmonic resonances can be ignored. So, henceforth we consider only
$\ell=1$. For small frequencies $\omega_R$, the turning point position
$z_A$ is approximately equal to the half-width of the hole, and varies
only logarithmically at low frequency, because it occurs at
$2\omega_R^2=-W=\phi_0(z)\simeq16\psi{\rm e}^{-|z|}$ in the wings, so
for an exact shift mode
\begin{equation}
  \label{eq:zA}
z_A
\simeq \ln\left(16\psi\over 2\omega_R^2\right)
=\ln\left(8\psi^{-1/2}\over (\omega_R^2/\psi^{3/2})\right)
\simeq \ln\left(8\psi^{-1/2}\over (0.05)^2\right)=8.-\ln(\psi)/2,
\end{equation}
using the observed numerical scaling $\omega_r/\psi^{3/4}\sim 0.05$.
The value $z_A=9$ will be used as a generic estimate.  The dominant
dependency is $\tF_R\propto \xi\omega_r^5/\psi^{1/2}$, and we will
ignore the weak dependence on $\psi$ and $\omega_r$ of $z_A$ and of the
correction factors $D$ and $A$, in estimating the constant of
proportionality.  For $\psi=0.16$, $D=0.76$, and $A$ can be as small
as $0.9$, in which case $D/A^4\simeq 1.15$ and the combined correction
factors are moderate giving
$\tF_R\simeq 1000 i\xi\omega_r^5/\sqrt{\psi}$.

\subsection{Non-resonant Force}
The non-resonant contribution to eq.\ (\ref{eq:phimint}) is the sum of
the non-resonant ($\int_0^t$) integral plus the term $i\omega z(t)$.
If treated using the square-wave approximation $z(t)=sign(t)z_A$ it is
\begin{equation}
  \label{eq:nonresphi}
{q_e\Phi_{NR}(t)\over i\omega\xi}=
z(t) +(i\omega)\int_{0}^tz(\tau) {\rm  e}^{-i\omega(\tau-t)}d\tau  
=sign(t)z_A{\rm e}^{i\omega t},
\end{equation}
taking into account that the signs of $z$ and $t$ are the same.
When
multiplied by the antisymmetric quantity $-(i\omega)^2q_e{d\phi_0\over dz}$ and
integrated for positive and negative $z(t)$ and $v$ this gives rise to
a force
\begin{equation}
  \label{eq:FNR}
  \tF_{NR} = -(i\omega)^2\xi \int_0^{\sqrt{-2q_e\psi}}{\partial f_0\over \partial W} 
  \int_0^{t_b/2} z_A2\cos\omega \tau' q_e{d\phi_0\over dz} d\tau'
  v_\psi dv_\psi.
\end{equation}
This expression is manifestly real when $\omega_i=0$, and provides the
main contribution to $\Re(\tF_t)$. But it also makes an important
contribution, first order in $\omega_i$, to $\Im(\tF_t)$: approximately
$(2\omega_i/\omega_r) \tF_{NR}(\omega_r)$.

It is less clear for the non-resonant contribution that the
square-wave approximation to $z(t)$ is accurate, since contributions
to this integral also arise from deeply trapped particles, which have
almost sinusoidal bounce motion.  Nevertheless, proceeding as before
to replace $-q_e{d\phi_0\over dz}={dv\over d\tau'}$, we get
$-\int_0^{t_b/2}\cos\omega\tau'q_e{d\phi\over
  dz}d\tau'=\int_0^{t_b/2}\cos\omega\tau'{dv\over d\tau'}d\tau' =
-v_\psi+v(t_b/2)\cos(\omega t_b/2)+\int \omega \sin\omega\tau'
v(\tau') d\tau'$.
The final integral term can be ignored by $\omega$ ordering and the
second because $v(t_b/2)=0$ everywhere except on a negligible measure
at $\omega t_b/2\simeq n \pi$; so we can substitute $ -v_\psi$  for
$-\int_0^{t_b/2}\cos\omega\tau'q_e{d\phi\over
  dz}d\tau'$, giving
\begin{eqnarray}
  \label{eq:FNR2}
  \tF_{NR} &\simeq& -(i\omega)^2\xi \int_0^{\sqrt{-2q_e\psi}}
  {\partial f_0\over \partial W}  z_A2v_\psi v_\psi dv_\psi\nonumber\\
  &\simeq& -(i\omega)^2\xi 
  \left\langle{\partial f_0\over \partial W}  2z_A\right\rangle 
  {1\over 3} (2\psi)^{3/2}= \omega^2\xi w_t
           f_{0s}\sqrt{2\pi}{5\over 8} \psi ,
\end{eqnarray}
where $\left\langle{\partial f_0\over \partial W} 2z_A\right\rangle$
denotes an average over energy whose weighting $\propto v_\psi^2$
(emphasizing \emph{shallowly} trapped orbits) justifies
writing it as $w_t f_{0s}(15/16)\sqrt{\pi/\psi}$, with
$w_t\sim 2z_A$ a hole width somewhat greater than unity. [But the
$z_A$ here is not just for resonant orbits; it is an average over all
trapped particles.]
This expression for unit density is $ {5\over 8} w_t \omega^2 \xi \psi
=5\omega^2\xi\psi$ when $w_t=8$ which is a plausible estimate.

The non-resonant $\omega_i$-term involves in addition a small
contribution from passing particles. The total value
$\Re(\tF/\omega^2)$ has been evaluated more precisely
elsewhere \citep{Hutchinson2016,Hutchinson2018a}, and for unit density
can be written in terms of the function
$ h(\chi) =-{2\over\sqrt{\pi}}\chi +(2\chi^2-1){\rm e}^{\chi^2}{\rm
  erfc}(\chi) +1=\chi^2-{2\over\sqrt{\pi}}{4\over 3}\chi^3+O(\chi^4),
$ where $\chi^2=\phi_0(z)$, as
\begin{equation}
  \label{eq:trapinertia}
  \tF_{NR} = \xi\omega^2\int h(\sqrt{\phi_0(z)}) dz=\xi\omega^2\psi w_i 
  \simeq \xi\omega^2\int\phi_0(z)dz,
\end{equation}
which can be considered to define a width $w_i\simeq\int\phi_0dz/\psi$.
For shallow holes of the chosen shape, it is
$w_i\simeq\int\sech^4(z/4) dz=16/3$, giving $\tF\simeq
5.3\xi\omega^2\psi$, in excellent agreement with the estimate
developed from eq.\ (\ref{eq:FNR2}), confirming that the passing
particle contribution is relatively unimportant here. 

\subsection{Passing-Particle Imaginary Force}

To obtain the lowest order imaginary part of the passing-particle
force that is independent of $\omega_i$ we proceed in a similar manner
using two integrations by parts, except with
different constants of integration than eq.\ (\ref{eq:phimint}).
\begin{equation}
  \label{eq:phipassint}
  {q_e\Phi(t)\over\xi}
  =[v(t)-v_\infty]+i\omega [z(t)-z_\infty(t)]
  +(i\omega)^2\int_{-\infty}^t[z(\tau)-z_\infty(\tau)] {\rm  e}^{-i\omega(\tau-t)}d\tau. 
\end{equation}
Here $v_\infty$ denotes the distant velocity (outside the hole) of the
orbit under consideration; and $z_\infty(t)=v_\infty t+const.$ denotes
the position as a function of time for motion at a constant velocity
$v_\infty$ which extrapolates the distant past orbit ignoring
$\phi(z)$. Thus $[z(\tau)-z_\infty(\tau)]$ is zero in the distant
past, rises as it passes through the hole, and then remains constant
past the hole. As before, it is the final integral term that gives the
imaginary force (real part of $\Phi$). We approximate the integral by
treating $[z(\tau)-z_\infty(\tau)]$ as the unit step function
$H(\tau)\equiv [1+sign(\tau)]/2$ times an amplitude
$z_p=\int_{-\infty}^\infty [v(t)-v_\infty] dt$, and note that
$\int_{-\infty}^tH(\tau){\rm e}^{-i\omega(\tau-t)}d\tau= H(t)[1-{\rm
  e}^{i\omega t}]/(-i\omega)$.
Then, ignoring other terms that are subsequently annihilated by
symmetry, we have for small $\omega t$ a term independent of
$\omega_i$
\begin{equation}
  \label{eq:rePhi}
  \Re\{q_e\Phi(t)\}\simeq\Re\{(-i\omega)\xi z_p[1-{\rm e}^{i\omega t}] H(t)\}
  \simeq (i\omega_r)^2 \xi z_p t H(t).
\end{equation}
From the second term of eq.\ (\ref{eq:phipassint}) we also have a
component of $q_e\Phi(t)$ equal to $i\omega_r z_pH(t)$, which could be
pursued, but it contributes imaginary force only $\propto\omega_i$
and is subdominant to the similar trapped component, so we do not
bother.  The required imaginary contribution to passing force at
$\omega_i=0$ is thus
\begin{eqnarray}\label{eq:impassing}
  i\Im(\tF_{p}) &\simeq& -\int \int q_e{d\phi_0\over dz}\xi  
                    (i\omega)^3{\partial f_0\over \partial W}
                    z_p \tau' H(\tau') d\tau' v_\infty dv_\infty\nonumber\\
  &=& -q_e\xi (i\omega)^3 \int{\partial f_0\over \partial W} z_p
      \int_0^\infty -\tau' {dv\over d\tau'} d\tau' v_\infty dv_\infty\nonumber\\
  &=&  - q_e\xi (i\omega)^3 \int{\partial f_0\over \partial W} z_p
      \int_0^\infty [v(\tau')-v_\infty] d\tau' v_\infty dv_\infty\nonumber\\
  &=& -q_e\xi (i\omega)^3 \int{\partial f_0\over \partial W}
      {z_p^2\over 2} v_\infty dv_\infty.
\end{eqnarray}
The integral quantity $z_p(v_\infty)=\int(1-|v_\infty/v|)dz$ is
approximately the value of its integrand at the hole center times the
hole width: $z_p(v_\infty)\simeq (1-v_\infty/\sqrt{v_\infty^2+2\psi})w_p $. Its
dependence for $v_\infty^2\gtrsim \psi$ is
$z_p\simeq w_p \psi/v_\infty^2$; so
$\int_\psi^\infty z_p^2dv_\infty^2\simeq w_p^2\psi$, and for small
$\psi$ we can multiply this by the value of $df_0/dW$ at small
energy. The remainder of the integral can be approximated as having constant
$z_p\simeq w_p$ so $\int_0^\psi z_p^2dv_\infty^2\simeq w_p^2\psi$,
giving
\begin{equation}
  \label{eq:vinfint}
   \int_0^\infty {\partial f_0\over \partial W}
      {z_p^2\over 4} dv_\infty^2 \simeq \left.{\partial f_0\over \partial
        W}\right|_{v_\infty=0} w_p^2 {\psi\over 2}
\end{equation}
(which can be considered the definition of $w_p$).
Thus finally 
\begin{equation}
  \label{eq:tFpi}
  \Im(\tF_{p})\simeq q_e\xi \omega^3 \left.\partial f_0\over \partial W\right|_0
  {w_p^2\over 2} \psi.
\end{equation}
This is negative because $\left.\partial f_0\over \partial
  W\right|_0=-f_{0s}=-1/\sqrt{2\pi}$ for unit density. Also notice
that despite depending on ${\partial f_0\over \partial W}$ this is not
a resonance effect but is contributed by all energies from zero to a few
times $\psi$.

\section{Solution of the Dispersion Relation}

\subsection{Complex Frequency}
We now have analytic approximations of the three terms that at low
frequency predominate the relation that the total imaginary part of
the hole force must vanish:
\begin{equation}
  \label{eq:disprel}
  2{\omega_i\over \omega_r} \Re(\tF) + \Im(\tF_t)+\Im(\tF_p)
  = \xi[C_i \omega_i\omega_r \psi + C_t \psi^{-1/2} \omega_r^5 + C_p \omega_r^3\psi]=0.
\end{equation}
Here the primary $\omega$ and $\psi$ dependences have been written explicitly,
leaving coefficients $C_i\simeq 10 $, 
$C_t\simeq 1000 $, and 
$C_p\simeq - w_p^2/2\sqrt{2\pi}\simeq -7$ (for $w_p=6$, a
plausible value chosen with hindsight) for unit density
$f_{0s}=1/\sqrt{2\pi}$ that are approximately constant. Writing
$\hat\omega_r=\omega_r/\psi^{3/4}$ and
$\hat\omega_i=\omega_i/\psi^{3/2}$ the relation becomes
\begin{equation}\label{eq:scaledrel}
  0= \Im(\tF)=\xi\psi^{13/4}\hat\omega_r
  [C_i\hat\omega_i +C_t\hat\omega_r^4+C_p\hat\omega_r^2]
\end{equation}
which demonstrates the universality (observed approximately in
Fig.\ref{fig:fcontours}) when expressed in the scaled quantites
$\hat\omega_r$ and $\hat\omega_i$. It also reproduces the quantitative
results of Fig. \ref{fig:fcontours}, and specifically the zero contour
of $\Im(\tF)$.
\begin{figure}
  \centering
 \ (a)\hskip-1.5em\includegraphics[width=0.47\hsize]{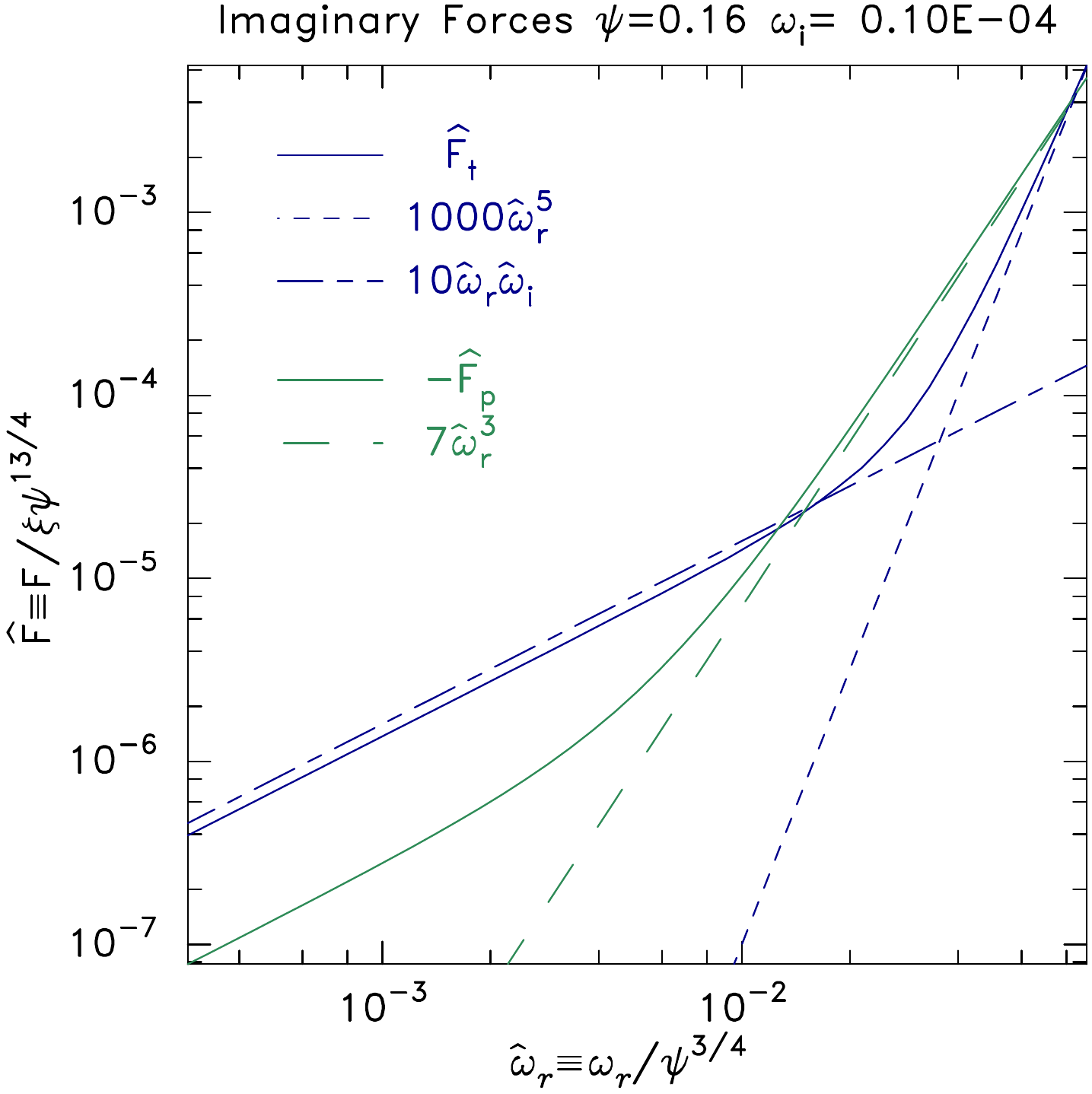}
 \ (b)\hskip-1.5em\includegraphics[width=0.48\hsize]{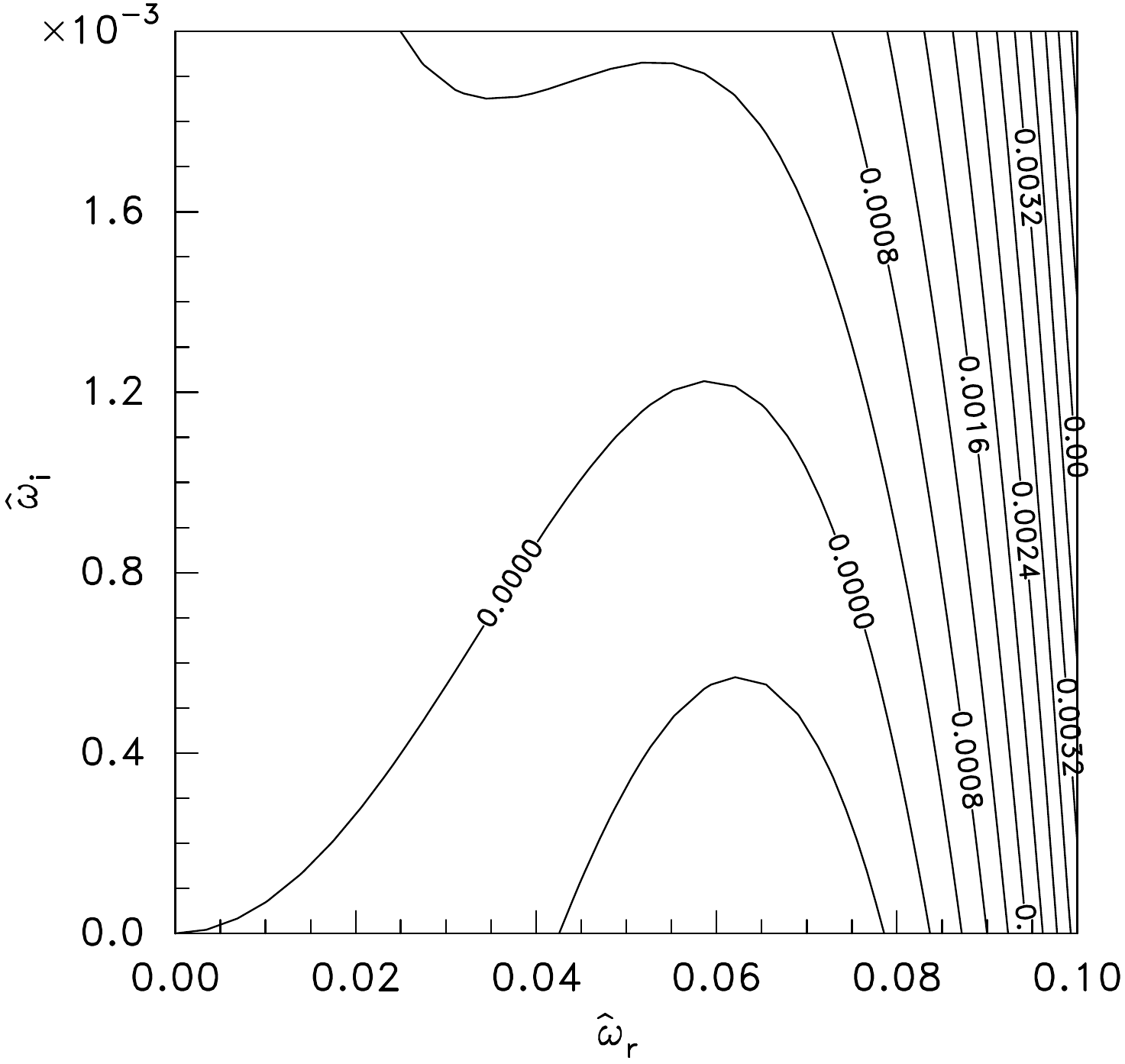}
  \caption{(a) Numerical integration (solid lines) agrees
    asymptotically with the analytic
    approximations (dashed lines) of the three terms in the imaginary force. The
    curve is universal when expressed in scaled quantities
    $\hat\omega_r$, $\hat\omega_i$ and $\hat F$. (b) The contours of
    force resulting from the analytic expressions, showing good shape
    agreement with Fig \ref{fig:fcontours}.}
  \label{fig:forces}
\end{figure}
First though, Fig. \ref{fig:forces}(a) demonstrates the agreement of
the analytic formulas for the three imaginary components of the force
with the corresponding parts of the force calculated by the numerical
integration code. All the forces in that plot are normalized in the
form $\hat F \equiv \tF/\xi\psi^{13/4}$, in accordance with
eq. (\ref{eq:scaledrel}), giving an essentially universal plot,
independent of $\psi$ (when small) expressed in terms of
$\hat\omega$. The agreement of the analytic curves with the
appropriate range of the full numerical integration is very good,
confirming that the dominant terms (of \ref{eq:disprel}) have been
correctly indentified and quantitatively estimated\footnote{The
  asymptotic low-frequency logarithmic slope of $\Im(\tF_p)$ is unity, the
  same as that of $\Im(\tF_t)$. It comes from the $z-z_\infty$ term of
  eq.\ (\ref{eq:phipassint}) whose coefficient we did not bother to
  estimate.}.

Fig.\ \ref{fig:forces}(b) shows the shape of the force contours for
comparison with Fig.\ \ref{fig:fcontours}. The agreement is within the
level of variation expected.
The position of the bottom right-hand end of the zero contour, at
$\omega_i=0$, is decided by setting equal and opposite the second and
third terms, whose solution can be rendered as
\begin{equation}
  \label{eq:omegarpsi}
  \hat\omega_r = \sqrt{-C_p/C_t}\simeq \sqrt{7/1000}=0.084.
\end{equation}
Near the bottom left hand end we can neglect the trapped term
and find
\begin{equation}
  \label{eq:omegaipsi}
  \hat\omega_i \simeq (-C_p/C_i)\hat\omega_r^2\simeq
  0.7\hat\omega_r^2. 
\end{equation}
This parabolic $\hat\omega_r$-dependence appears to be present in
Fig.\ \ref{fig:fcontours} (though numerical rounding and other
uncertainties make the zero contour shape imprecise) and is obvious in
Fig.\ \ref{fig:forces}(b).  The top of the
contour is determined by maximizing
$-C_t\hat\omega_r^4-C_p\hat\omega_r^2$ which gives
\begin{equation}
  \label{eq:tripos}
\hat\omega_r=\sqrt{-C_p\over 2C_t}
\simeq 0.06
\end{equation}
and then
\begin{equation}
  \label{eq:triheight}
  \hat\omega_i =  {1\over 2}\left(-C_p\over C_i\right) \omega_r^2
  = {1\over 4}\left(-C_p\over C_i\right)
  \left(-C_p\over  C_t\right)
  \simeq 1.2\times10^{-3}.
\end{equation}
These expressions agree with the corresponding positions on the
zero contour in Fig. \ref{fig:forces}(b) (as they must).

It is clear from these observations that the effect of the resonant
force is to \emph{limit} the maximum $\omega_r$ at which instability
can occur. In other words, it is a stabilizing term (contradicting
Refs.\ \citep{Newman2001a,Vetoulis2001,Berthomier2002}), which as
$\omega_r$ increases eventually overcomes the others because it varies
as a higher power. If its coefficient $C_t$ were reduced, for example
by decreasing the distribution gradient
${\partial f_0\over \partial W}$ for shallowly trapped particles, or
for a perturbation whose shift was smaller in the wings of the hole,
the effect would be to permit instability to higher $\omega_r$ and
with higher $\omega_i$, following the
$\hat\omega_i\simeq0.35\hat\omega_r^2$ trajectory further up.  The
stabilizing nature of the resonance might appear counter-intuitive
since ${\partial f_0\over\partial W}$ is positive causing resonant
particles to give energy to the eigenmode. What makes this effect
stabilizing, however, is that the eigenmode is a ``negative energy''
mode, reflecting the negative inertia of the hole. Therefore adding
energy to it tends to \emph{reduce} its amplitude.

\subsection{Wavenumber} 

As explained earlier, the wavenumber must be chosen to satisfy the
real part of eq.\ (\ref{eq:forcebalance}) $\Re(\tF)=F_E$. The real
frequency $\omega_r$ gives the predominant real part of $\tF$ at low
frequency, namely $\Re(\tF)=(C_i/2)\xi\omega_r^2\psi$, see eqs.\
(\ref{eq:trapinertia}) and (\ref{eq:disprel}). The Maxwell stress force can
be evaluated in dimensionless units as
\begin{equation}
  \label{eq:elecforce}
  F_E=\xi k^2\int\left(d\phi_0\over dz\right)^2dz={128\over 315}\xi
  k^2\psi^2, 
\end{equation}
from which we conclude
\begin{equation}
  \label{eq:kvalue}
  k^2={315\over 128}{C_i\over2}{\omega_r^2\over \psi}, \qquad {\rm
    giving}\quad \hat k ={k\over\psi^{1/4}} = \sqrt{315 C_i\over
    256}\hat\omega_r =3.5\hat\omega_r.
\end{equation}
At the maximum unstable $\omega_i$, $\hat\omega_r\simeq 0.06$, we then
require $\hat k \simeq 0.2$, which is in reasonable agreement with
Fig. \ref{fig:fcontours}. We have recovered the observed
scaling of the universal solution: $k= \hat k \psi^{1/4}$, and
approximately the absolute value of $\hat k$ required for maximum
instability growth.

\section{Comparison with simulation}

The high-B ($\Omega>2\omega_p$) instability investigated here was
first observed in particle in cell simulations of initially uniform
two-stream electron
distributions \citep{Goldman1999,Oppenheim1999,Oppenheim2001b}, and has
been observed in similar simulations
since \citep{Umeda2006,Lu2008,Umeda2008}. The electron holes form from
non-linear Langmuir instability and then usually experience transverse
instabilities coupled to external waves. Umeda et al \citep{Umeda2006}
observed that warm beams with significant velocity spread give
shallower holes, and they found no transverse hole instability for
depth $\psi\lesssim 0.8$ at high-B.  Such treatments may well be
appropriate for a full scale simulation of events of nature, but for
insight into the instability mechanisms, they suffer from difficulties
associated with the highly distorted and uncertain electron velocity
distributions and a high level of fluctuations associated with
different wave phenomena.

Single holes deliberately set up by initial simulation conditions have
been studied as an alternative that provides cleaner insight in the
linear instability growth.  Oppenheim et al \citep{Oppenheim1999}
generated slab holes from (effectively) one-dimensional two-stream
simulations, which were then used as initial conditions for a
two-dimensional PIC simulation. It is not clear what the distribution
function $f_0$ actually was when multidimensional dynamics was turned
on. Publications observing high-B hole-wave instability of pre-formed holes
of specified distribution seem to be limited to the continuum Vlasov
simulation of deep waterbag holes by Newman et al \citep{Newman2001a}
and the PIC simulations of Wu et al \citep{Wu2010} ($\psi=0.8, 2$). A
waterbag distribution, for which $\partial f_0/\partial W=0$,
unfortunately makes zero (or, at the waterbag boundary, infinite) the
crucial resonant term. Wu's potential profiles were Gaussian,
corresponding still to rather pathological electron velocity
distributions with infinite gradient at the separatrix; and they
documented mostly the nonlinear phases.

Therefore new simulations have been run to compare with the present
theory. They use the COPTIC particle in cell
code \citep{Hutchinson2011a}\footnote{Available from
  https://github.com/ihutch/COPTIC}, pushing only electrons (typically
$\sim1$ billion, on 512 processors) on a 2-D rectangular periodic
domain with mesh spacing 1 (Debyelength) and timestep 0.5. They have
effectively infinite magnetization. A hole of specified depth $\psi$
is created at the center $z=0$ by prescribing the initial particle
distribution solved for by the integral equation BGK method, and
loaded using a quiet-start particle placement. The initial hole has a
modified potential form
$\phi=\psi[1+\exp(1/\psi)]/[1+\exp(1/\psi)\cosh^4 z/4]$ which has a
stretched potential top unless $\psi$ is rather small. This
modification recognizes that an exact $\sech^4z/4$ hole cannot exist
for $\psi>0.75$ because it requires a negative central distribution
function $f_0<0$ at $W=-\psi$. The potential modification causes a
modest (typically $<30$\%) change in $\partial f_0/\partial W$, which
is smaller than other simulation differences discussed later. A
summary of the cases run is given in table \ref{tab:runs}.

\begin{table}
  \centering
  \begin{tabular}{ccccccccccccc}
    $\psi$&$nz$&$ny$&Time&Mode&$A_{max}$&${2\pi\over\omega_r}$&${1\over\omega_i}$
    &$\hat\omega_r$&$\hat\omega_i$
    &$\hat\omega_i/\hat\omega_r^2$&$\hat k$&$\hat k/\hat\omega_r$\\
1	&1024	&256	&3200	&8\dag	&22	&92	&522	&0.0683	&0.0019	&0.41	&0.20	&2.87\\
1	&512	&200	&2500	&6	&25	&96	&477	&0.0654	&0.0021	&0.49	&0.19	&2.88\\
1	&512	&212	&3000	&10\dag	&8	&65	&391	&0.0967	&0.0026	&0.27	&0.30	&3.07\\
1	&512	&224	&2500	&7	&25	&94	&340	&0.0668	&0.0029	&0.66	&0.20	&2.94\\
1	&512	&238	&3000	&7	&5	&94	&580	&0.0668	&0.0017	&0.39	&0.18	&2.76\\
1	&512	&256	&2500	&8	&25	&94	&375	&0.0668	&0.0027	&0.60	&0.20	&2.94\\
0.8	&600	&200	&3000	&6	&20	&110	&440	&0.0675	&0.0032	&0.70	&0.20	&2.95\\
0.8	&512	&200	&2700	&1	&1	&	&	&	&	&	&	&\\
0.8	&512	&212	&2000	&7	&4	&100	&460	&0.0743	&0.0030	&0.55	&0.22	&2.95\\
0.8	&512	&224	&3000	&8	&6	&105	&455	&0.0707	&0.0031	&0.61	&0.24	&3.35\\
0.8	&512	&238	&2600	&8	&19	&100	&443	&0.0743	&0.0032	&0.57	&0.22	&3.01\\
0.8	&512	&256	&3000	&1	&1.5	&	&	&	&	&	&	&\\
0.6	&600	&200	&3000	&3	&1.2	&	&	&	&	&	&	&\\
0.6	&550	&200	&3000	&7	&9	&110	&560	&0.0838	&0.0038	&0.55	&0.25	&2.98\\
0.6	&532	&200	&2600	&7	&12	&110	&460	&0.0838	&0.0047	&0.67	&0.25	&2.98\\
0.6	&512	&200	&3000	&7	&8	&106	&513	&0.0869	&0.0042	&0.55	&0.25	&2.87\\
0.6	&512	&212	&3000	&2	&1.2	&	&	&	&	&	&	&\\
0.6	&512	&224	&3000	&8	&6	&107	&480	&0.0861	&0.0045	&0.60	&0.25	&2.96\\
0.6	&512	&238	&3000	&2	&0.8	&	&	&	&	&	&	&\\
0.6	&512	&256	&3000	&9	&10	&106	&490	&0.0869	&0.0044	&0.58	&0.25	&2.89\\
0.4	&512	&200	&5200	&8	&9	&124	&715	&0.1007	&0.0055	&0.54	&0.32	&3.14\\
0.4	&512	&212	&3000	&1	&0.8	&	&	&	&	&	&	&\\
0.4	&512	&224	&4600	&9	&9	&120	&805	&0.1041	&0.0049	&0.45	&0.32	&3.05\\
0.4	&512	&238	&3000	&2	&0.8	&	&	&	&	&	&	&\\
0.4	&512	&256	&5000	&10	&1	&118	&	&0.1059	&	&	&0.31	&2.92\\
0.3	&512	&200	&8000	&3	&1.2	&	&	&	&	&	&	&\\
0.3	&512	&212	&10000	&1	&1.2	&	&	&	&	&	&	&\\
0.2	&512	&200	&3000	&2	&0.5	&	&	&	&	&	&	&\\
0.2	&512	&212	&3000	&1	&0.7	&	&	&	&	&	&	&\\
  \end{tabular}
  \caption{Summary of the PIC hole instability simulations. See text
    for detailed explanation.}
  \label{tab:runs}
\end{table}
It is observed (significantly, and in agreement with earlier reports)
that the stability outcome is affected somewhat unpredictably by small
changes in the size of the mesh, $nz$, $ny$, which is why multiple
domain size cases are run for each hole depth. This appears to be a
physics effect arising from the finite periodic domain. It shows that
the hole ``knows about'' conditions a long way away, and the
dependence on $z$-extent implies that coupling to external waves is a
significant influence. The two cases where the unstable mode number is
labelled with a dagger observe two periods of the low-level waves
spanning the $z$-domain, all others have one period.

\begin{figure}
  \centering
  \includegraphics[width=0.8\hsize]{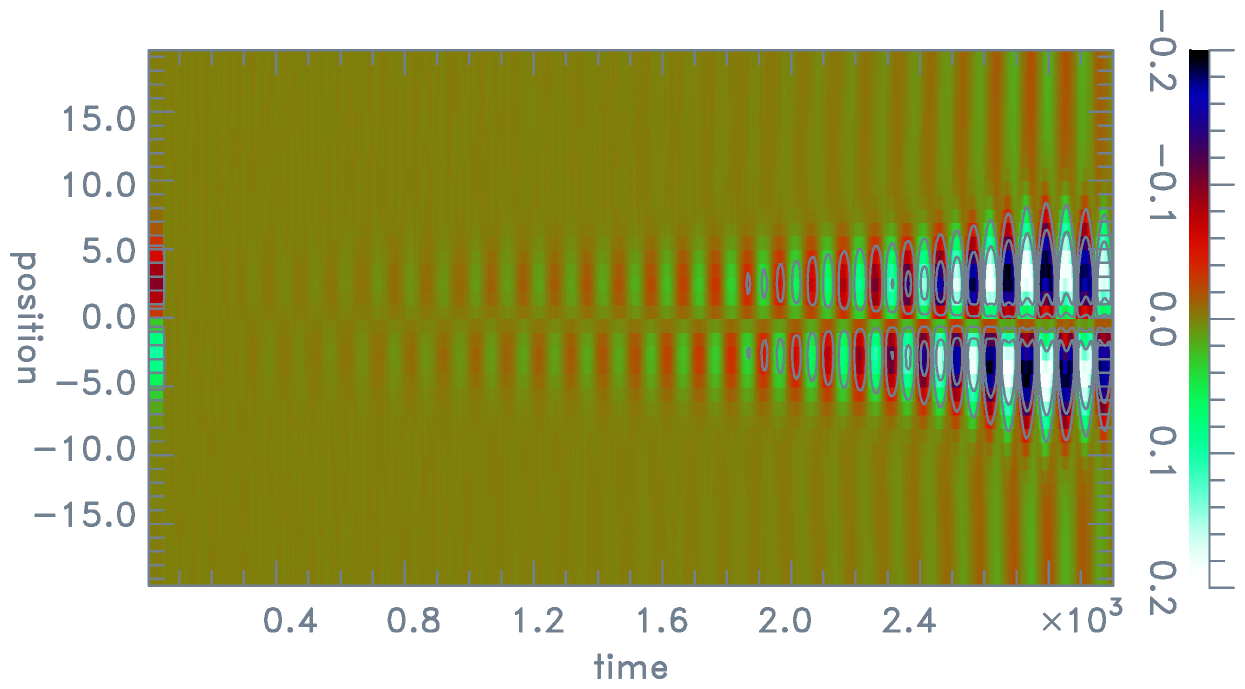}
  \caption{Color (shaded) contours over the indicated range of the
    oscillating mode ($j=$)8 amplitude $A_8(t,z)$ for case $\psi=0.8$,
    $nz=512$, $ny=238$, showing its growth and spatial variation over
    the central region $|z|\le 20$. }
  \label{fig:modegrowth}
\end{figure}

A useful diagnostic of instability is provided by Fourier-transforming
the potential in the transverse $y$-direction and examining the time
and ($z$-)space behavior of the lower order ($j=0-10$) mode
amplitudes:
$A_j(t,z)=\Re[{1\over ny}\sum_{p=0}^{ny-1}\phi(p,z,t)\exp(-i2\pi
pj/ny)]$,
where $p$ is the $y$-position index. This has much greater sensitivity
than merely inspecting the spatial distribution of potential or
density, because it appropriately averages over the mode structure. It
also lends itself to a simple contour plot display of amplitude on the
$z$-$t$ domain which documents the eigenmode structure and
evolution. Unstable cases observe certain modes growing from
noise-level fluctuations. Sometimes, even early on when mode
amplitudes are small, different modes grow and saturate at a low level
or decay away, but for $\psi\ge 0.4$, usually a particular mode
eventually begins to dominate and grows exponentially until it becomes
nonlinear. That mode number (the number of wavelengths spanning the
$y$-domain) is noted in the table.  Fig \ref{fig:modegrowth} shows an
example of an unstable mode's time development.
In the table the $A_{max}$ column gives the mode's maximum amplitude
(times 100 for compactness); and the `time' column refers to the time the
growing mode is largest, which for large amplitudes ($\gtrsim 10$) is
before the end of the actual run when nonlinear behavior sets in, but
for small or non-growing cases is the run duration (and Mode and
$A_{max}$ then refer to the largest amplitude during the run). 

The growing mode illustrated in Fig.\ \ref{fig:modegrowth} saturates at time
approximately 2600. An arbitrarily scaled template exactly equal to
the shift mode replaces the first 20 time slots of the contour
plot. The shape of the later actual local perturbation is rather
similar to it, but is accompanied by a synchronized wave component
consisting of a low-amplitude potential perturbation almost uniform in
space beyond $|z|\gtrsim 10$.  Prior to saturation there appears to be
approximately a 90 degree phase shift between the wave and the shift
perturbation, but as saturation occurs the wave becomes closer to 180
degrees out of phase with the shift.

For shallower holes, $\psi< 0.4$, no long-term exponentiation was
definitively observed, even in cases run for as long as $t=10000$. The
very strong scaling of the hole force $\tF \sim \psi^{13/4}$ and the
lower growth rates means lower $\psi$ cases are more affected by noise
or other simulation inaccuracies. However, the high
computational cost of long duration, low-$\psi$, simulations
discouraged more thorough investigation, and it cannot definitively be
concluded that there is never instability for $\psi< 0.4$.

The oscillatory period (${2\pi\over\omega_r}$) and growth time
($1\over \omega_i$) of clear instabilities, when present, are
documented in the table. Their uncertainty is conservatively estimated
at $\pm10$\% and $\pm20$\% respectively. Several of the scaled forms
of these parameters are also given. We observe that $\hat\omega_r$
increases as the hole depth is decreased, to as much as a factor of 2
above what is predicted by the analysis; and $\hat\omega_i$ to as much
as a factor of 4; although the ratio $\hat\omega_i/\hat\omega_r^2$
remains constant as predicted. The ratio $\hat k/\hat\omega_r$, which
according to analysis depends only on the real part of the force,
varies little, lying typically only 15\% below the value 3.5 the
analysis predicts. Thus, we have good confirmation of the real part of
the force analysis.

The instability occurring at somewhat higher $\omega$ than predicted
by analysis would be expected if the resonant stabilizing force is
weaker, because the effective $\partial f_0/\partial W|_R$ is
reduced. Sorting particles in the simulation by total energy
($W=-\phi+v^2/2$) shows that the noise level of $\phi$ is sufficient
to broaden the apparent energy distribution by an amount of order 0.01
($T_e$ units), presumably flattening the effective $f_0$ near its
slope discontinuity at $W=0$. The resonant energy is
$-W\simeq2\omega_r^2=2\hat\omega_r^2\psi^{3/2}$ which is of order
0.004 or less. Therefore the $\phi$ noise is clearly capable of a
strong influence on the relevant resonant part of the distribution
function, and would be expected to reduce the slope
$\partial f_0/\partial W|_R$. Perhaps it does so by the factor of 4
required to explain the $\omega_i$ enhancement, but it does not seem
possible to prove this hypothesis definitively. The resonance physics
is clearly very hard to reproduce and document in a simulation because
of the smallness of the resonant energy, and the slope discontinuity
at $W=0$.  Alternatively the stabilizing force might be weakened by
differences of the eigenmode from the pure shift, for example as a
result of the coupled wave component. Detailed consideration of such
effects is beyond the present scope.

\section{Summary}

Theoretical analysis shows that there is a shift-mode kink instability,
having low real frequency and very low growth rate, of an initially
planar electron hole at high magnetization ($\Omega\gtrsim 5\omega_p$
in dimensional units). For a hole of potential form
$\phi=\psi\sech^4(z/4\lambda_D)$, the predicted values of the fastest
growing mode are $\omega_r/\omega_p\simeq 0.06(e\psi/T_e)^{3/4}$,
$\omega_i/\omega_p\simeq 0.001(e\psi/T_e)^{3/2}$, and
$k\lambda_D= 3.5 \omega_r/\omega_p (e\psi/T_e)\simeq0.2(e\psi/T_e)^{1/4}$.

A simple but accurate new approximation
$\omega_b/\omega_p=\sqrt{-W/2T_e}$ for the dependence of bounce
frequency $\omega_b$ on trapped particle energy $ W$ (for small $|W|$)
undergirds these purely analytic results. Numerical evaluation of the
solution of the Vlasov equation confirms the resulting identification
of the three predominant force terms giving rise to the instability
as: (1) the intrinsically imaginary part of the jetting of passing
particles (which is destabilizing), (2) the imaginary part of the
trapped particle resonant term (which is stabilizing), balanced
against (3) the imaginary part of the frequency affecting the (otherwise)
real total jetting.

Particle in cell simulations show good agreement with the predicted
$\omega_r$ and $k$, but faster growth by factors of 2 - 4. This
discrepancy appears explicable by non-ideal effects in the simulation
affecting the trapped distribution function, because the resonant energy is
extremely close to zero. Proximity to the distribution slope
theoretical discontinuity at the separatrix can be expected to
substantially reduce the effective positive slope of the trapped
distribution, \emph{reducing} stabilization.  The observed sensitivity
of the simulation results to small changes in domain size suggests
that hole coupling to the observed long wavelength external waves is
significant in them. However, the instability mechanism analyzed does
not depend on that coupling. The interpretation proposed is that the
hole-wave coupling is a sub-dominant effect of an instability
occurring in the hole itself. In that case, the instability can be expected to
occur in space plasmas whose domains do not impose the restrictive
boundaries and resulting resonant parallel wavelengths of simulations.

The strong reduction in growth rate as the hole depth ($\psi$) is
reduced predicts that shallow holes can be very long lived
indeed. Moreover if a hole has transverse extent $L_y$ insufficient to
accommodate the highest unstable wavenumber
$k_{max}\simeq 0.3(e\psi/T_e)^{1/4}/\lambda_D$, because
$k_{max}L_y<2\pi$, then presumably it will
be fully stable.

\section{Acknowledgements}

  I am grateful to Xiang Chen for pointing out the closed form
  solution eq.\ (\ref{eq:f0sech4}) for the distribution function.
  This work was partially supported by NASA grant NNX16AG82G. PIC
  simulations were carried out on the MIT-PSFC partition of the
  Engaging cluster at the MGHPCC facility (www.mghpcc.org) which was
  funded by DoE grant number DE-FG02-91-ER54109.

\bibliography{MyAll}

\end{document}